\documentclass{mem}
\usepackage{natbib}\usepackage{txfonts}\usepackage{balance}
\usepackage{graphicx}
\usepackage[a4paper,breaklinks,dvipdfm]{hyperref}
\idline{75}{282}
\begin{document}
\def\teff{$T\rm_{eff }$}
\def\kms{$\mathrm {km s}^{-1}$}

\title{
Red supergiant star studies with CO$^5$BOLD and Optim3D
}

   \subtitle{}

\author{
B. Plez\inst{1} 
\and A. Chiavassa\inst{2}
          }

  \offprints{B. Plez}

\institute{
Laboratoire Univers et Particules de Montpellier,
cc\,072, CNRS, 
Universit\'e Montpellier 2,
34095 Montpellier cedex 5,
France
\email{bertrand.plez@um2.fr}
\and
Laboratoire Lagrange,
Universit\'e de Nice Sophia-Antipolis,
CNRS,
Observatoire de la C\^ote d'Azur, 
Nice, France
}

\authorrunning{Plez \& Chiavassa}

\titlerunning{Red Supergiants}

\abstract{We describe recent work focused towards a better understanding of red supergiant stars using 3D radiative-hydrodynamics (RHD) simulations with CO$^5$BOLD. A small number of simulations now exist that span up to seven years of stellar time, at various numerical resolutions.  Our discussion concentrates on interferometric and spectroscopic observations. We point out a number of problems, in particular the line depth and line width that are not well  reproduced by simulations. The most recent introduction of a non-grey treatment of the radiation field dramatically improved the match with observations, without solving all difficulties. We also review the newly revived effective temperature scale controversy, and argue that it will only be solved using 3D RHD models.

}
\maketitle{}

\section{Red supergiant stars}
Red supergiant (RSG) stars are the largest stars in the universe and are among the brightest in the optical and
near-infrared. These stars with masses between roughly 10 and 40~$M_{\odot}$ have luminosities of 20\,000 to 300\,000\,$L_\odot$,
     and radii in excess of 1\,500\,$R_\odot$ \citep{2005ApJ...628..973L}. Their effective temperatures, \teff, seemingly range from about
     3\,400 to 4\,100\,K. RSGs exhibit variations of integrated brightness,
of surface features, and of spectral line depth, shape, and position \citep{2007A&A...469..671J,2008AJ....135.1450G}. 
We also discuss below newly uncovered difficulties concerning their \teff determination.
As a consequence, RSG fundamental  parameters are difficult to determine, and their chemical composition remains largely unknown, despite the work of 
 e.g. \citet{2000ApJ...530..307C}, and \citet{2007ApJ...669.1011C}. In this paper we concentrate on the impact of CO$^5$BOLD simulations on the study of RSGs, mostly discussing high-angular resolution and spectroscopic observations.

\section{CO$^5$BOLD simulations of red supergiants}
Details on the radiation-magneto-hydrodynamics CO$^5$BOLD code can be found in \citet{2012JCoPh.231..919F, 2008A&A...483..571F, 2010ascl.soft11014F}, and elsewhere in  this volume. CO$^5$BOLD is designed to account for the physical conditions found in the atmosphere of cool stars and substellar objects, as well as in the underlying convective layers.
For the computation of RSG stars, we use the star-in-a-box setup: the computational domain is a cubic grid with a constant grid cell size. An open boundary condition is employed for all sides of the computational box.  The inner boundary condition is prescribed within a small spherical region in the centre of the cube where an internal energy source term provides the stellar luminosity. The largest RSG models computed so far contain 401$^3$ cells with a size of  2.5\,${\rm R_{\odot}}$ (grey), and  235$^3$ cells of size 8.6\,${\rm R_{\odot}}$ (non grey). The non-grey simulation was calculated with five opacity bins (\cite{1994A&A...284..105L}, and Ludwig, this volume). Both these models are described in \citet{2011A&A...535A..22C}.
The opacities used in the RSG models come from two different sources merged around 12\,000\,K. The low-temperature data is from PHOENIX \citep{1997ApJ...483..390H}, and the high-temperature from OPAL \citep{1992ApJ...397..717I}.
The relaxed RSG simulations span of the order of 3 to 7 years of stellar time.

\section{Optim3D LTE radiative transfer}

Although CO$^5$BOLD outputs bolometric intensity maps, it is necessary to compute detailed wavelength dependent intensities if one is to compare the simulations with observations. We developed the Optim3D LTE radiative transfer code for this purpose \citep{2009A&A...506.1351C}. 
In short, Optim3D allows the computation of detailed spectra emerging from each cell of the simulation box, accounting for the velocity field. The required computing resources are kept  to a minimum through the use of an efficient Laguerre-polynomial quadrature scheme for the intensity, and of a pre-tabulation of the monochromatic opacities with the Turbospectrum code \citep{Alvarez1998, 2012ascl.soft05004P}. Limitations to be lifted in the future are the impossibility to treat scattering, or the absence of dust opacities.
A number of recent papers illustrate the capabilities and impact of the CO$^5$BOLD and Optim3D pair, in imaging and interferometry \citep{2009A&A...506.1351C, 2010A&A...515A..12C,2010A&A...511A..51C}, in variability and astrometry studies, esp. related to Gaia \citep{2011A&A...528A.120C}, in spectroscopy \citep{2011A&A...535A..22C}, or in the study of the  atmospheric dynamics of RSGs \citep{2011A&A...528A.120C}.

\section{Computation of images and interferometric observations}

\begin{figure}[]
\resizebox{\hsize}{!}{\includegraphics[clip=true]{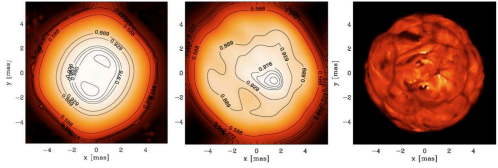}}
\caption{Reconstucted image from VLTI-AMBER observations of VX Sgr (left panel) compared to the best visiblity- and phase-matching 3D image from a CO$^5$BOLD simulation (right panel), and the same synthetic image convolved with the PSF of the instrument (central panel). From \citet{2010A&A...511A..51C}
\footnotesize
}
\label{fig-image}
\end{figure}

Interferometric observations potentially carry a great deal of information on stellar surface structures. Until recently most such observations were restricted to few telescopes and baselines, with the consequence of a limited amount of information being extracted. Typically a fit through a few points in the first or second lobe of the visibility would provide a uniform, or at best a limb-darkened disk diameter. Large modern interferometers operating with 3, 4 or more telescopes, and baselines that can be reconfigured, deliver now detailed visibility and phase closure information. RSG stars were observed in increasingly refined detail by several groups in the past two decades.  \citet{1990MNRAS.245P...7B,1992MNRAS.257..369W,1997MNRAS.285..529T,1997MNRAS.291..819W, 2000MNRAS.315..635Y} detected time-variable structures on the surface of Betelgeuse using the WHT, and  COAST.
\citet{2009A&A...508..923H} reconstructed an image of Betelgeuse in the H band, with two spots. \citet{2009A&A...503..183O,2011A&A...529A.163O} may have detected convective motions of the CO line forming layers.
\citet{2009A&A...506.1351C, 2010A&A...515A..12C} could confirm the presence of large-scale granulation on Betelgeuse by fitting visibility and phase closures computed from CO$^5$BOLD RSG simulations. Images of VX\,Sgr were reconstructed, although still with a low resolution \citep{2010A&A...511A..51C}, as shown on Fig.~\ref{fig-image}. The papers by \citet{2010SPIE.7734E.106C} and \citet{2012A&ARv..20...53B} show what can  typically be achieved today or soon will be, also in other fields than RSG studies. 

A difficulty with interferometers is the limited dynamic range of fringe detection. This sets stringent limits on the (resolution $\times$ field of view) product : observing small details on a big star is very difficult. 
Nevertheless, as outlined above, great progress was made recently in the interpretation of interferometric data. CO$^5$BOLD simulations have been massively used to generate model intensity distributions, then convolved with the spatial and spectral resolution of the instruments. 
Least-square fits of visibility and closure-phase data with simulation snapshot time-series have been used to find the best fitting model. Although the number of available simulations is still scarce, this allows a much better characterization of asymmetries due to granulation on real stars than the simple spot model.

\begin{figure}[]
\resizebox{\hsize}{!}{\includegraphics[clip=true]{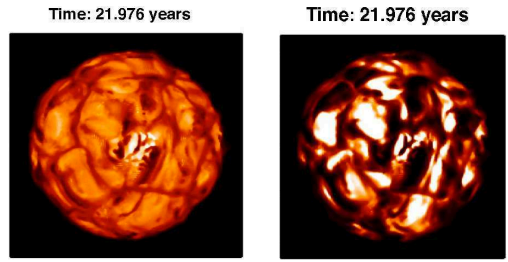}}
\caption{
\footnotesize
Images computed with Optim3D from the same snapshot of an RSG CO$^5$BOLD simulation; Left panel is in the K band with mostly contribution from the continuum \citep{2009A&A...506.1351C}, and right panel is in the optical dominated by strong TiO absorption \citep{2011A&A...528A.120C}.}
\label{fig-IR-vis}
\end{figure}

Calculations made by \citet{2011A&A...535A..22C} show the large difference in appearance of RSG simulations when seen at optical and IR wavelengths (Fig.~\ref{fig-IR-vis}). In the optical, TiO line absorption reveals highly contrasted structures smaller than the large granules seen in the IR. This is due to the strong temperature dependence of the source function at shorter wavelengths. Although observations in the optical are more challenging, and only one instrument, the Mount Wilson CHARA interferometer, has this capability \citep{2005ApJ...628..453T,2009A&A...508.1073M}, this opens very exciting prospects. 

\section{Computation of spectra}

\begin{figure}[]
\resizebox{\hsize}{!}{\includegraphics[clip=true]{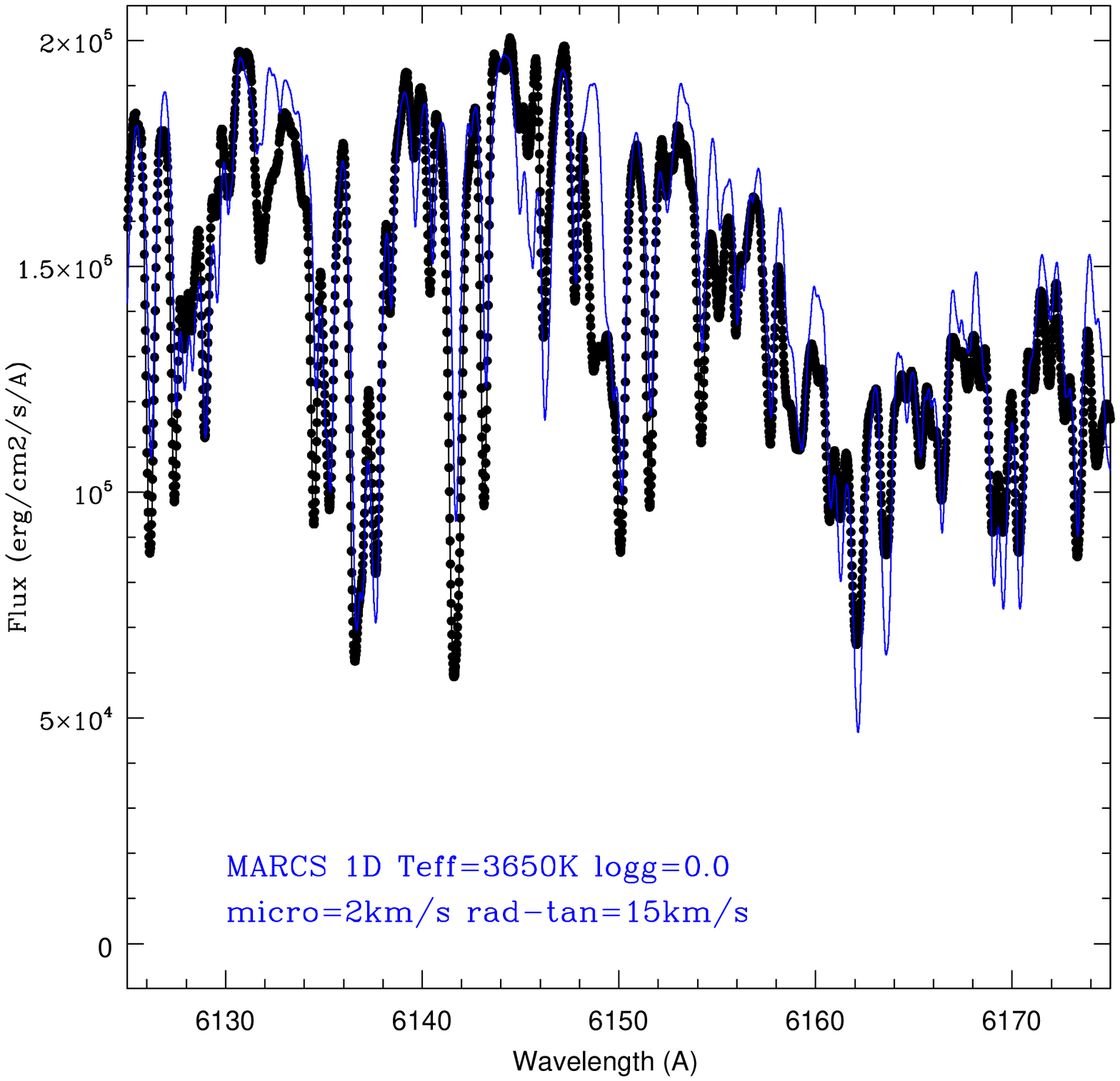}}
\caption{
\footnotesize
Synthetic spectrum from a 1D MARCS model with \teff=3650K (blue line) around a TiO band head at 6160\AA, and corresponding observed spectrum of Betelegeuse (black line and dots) from the ESO POP-UVES library  \citep{2003Msngr.114...10B} 
}
\label{fig-spec1D}
\end{figure}

\begin{figure}[]
\resizebox{\hsize}{!}{\includegraphics[clip=true]{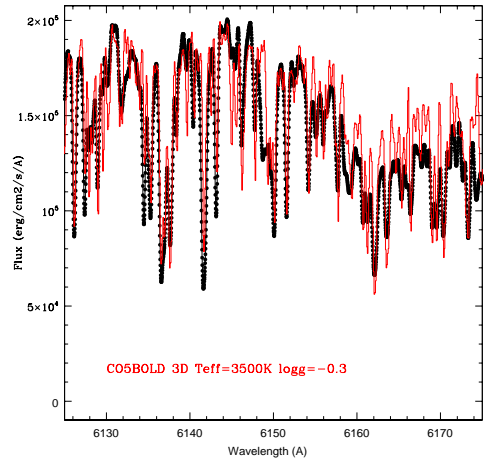}}
\caption{
\footnotesize
The same Betelgeuse spectrum with an Optim3D synthetic spectrum from a snapshot of a grey CO$^5$BOLD  simulation with \teff=3500K (red line). Despite the lower \teff, the TiO band head around 6160\AA\ appears too shallow.
}
\label{fig-spec3D}
\end{figure}

\begin{figure}[]
\resizebox{\hsize}{!}{\includegraphics[clip=true]{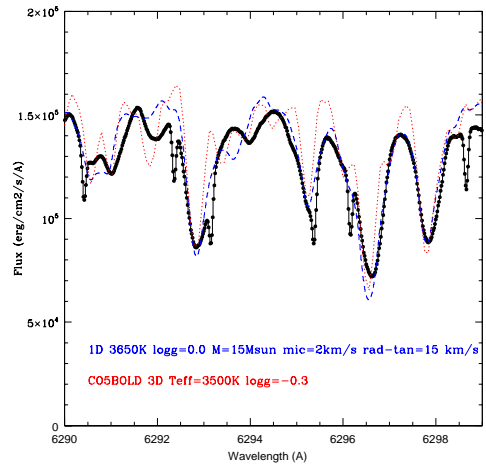}}
\caption{
\footnotesize
The same Betelgeuse spectrum at a slightly longer wavelength. The narrow glitches are telluric O$_2$ lines. The MARCS 1D spectrum (blue dashed line) was computed with a micro-turbulence parameter of 2\,km/s, and convolved with a radial-tangential profile \citep{2008oasp.book.....G} of 15\,km/s width. It fits the observed spectrum rather well. The Optim3D synthetic spectrum (red dotted line) is computed without any other velocity field than the one predicted in the CO$^5$BOLD simulation. Line profiles are too narrow.
}
\label{fig-spec3Dzoom}
\end{figure}

Being in Heidelberg, where spectroscopy was turned into an analysis tool by G. Kirchhoff and R. Bunsen around 1860, we are bound to discuss spectra emerging from CO$^5$BOLD simulations of RSGs. The first spectra computed with Optim3D did not reproduce real RSG spectra. They failed in two respects : (i) spectral features were not strong enough, esp. TiO bands appeared too shallow, and (ii) despite the presence of a supersonic velocity field, line profiles were too narrow. In an attempt to resolve the first issue, and noting that TiO bands become stronger with lower \teff\ in 1D models, we strived to run CO$^5$BOLD simulations with lower temperatures, and larger radii. This turned out to be very tricky and often lead to crashes. The coolest grey RSG simulation we could stabilize has \teff\ = 3487\,K, 830\,R$_{\odot}$, and 12\,M$_{\odot}$ (st35gm03n07, with 235$^3$ cells, in \cite{2011A&A...535A..22C}). This is at the lore end of the temperatures derived from fits of optical spectrophotometry  with 1D model spectra \citep{2005ApJ...628..973L, 2006ApJ...645.1102L, 2009ApJ...703..420M}. Still, the calculated TiO bands are too weak in this model (Fig.~\ref{fig-spec1D}, Fig.~\ref{fig-spec3D}). We speculated that a better, non-grey, treatment of the radiative field could solve the problem, as a better cooling in the optically thin layers would bring the temperature gradient closer to radiative equilibrium. 
Indeed, the inclusion of non-grey opacities with five bins in one simulation \citep{2011A&A...535A..22C} resulted in important changes in the atmospheric thermal structure (st35gm03n13, 235$^3$, with \teff\ = 3430\,K, 846\,R$_{\odot}$, and 12\,M$_{\odot}$). The average $<$3D$>$ temperature structure is more similar to a 1D hydrostatic model, because of more efficient radiative cooling in the optically thin layers. Temperature fluctuations at a given optical depth are also smaller in the non-grey simulation.
This single non-grey simulation available to us turns out to partially solve the line depth problem. The TiO band strengths are comparable to a 1D MARCS \citep{2008A&A...486..951G} model with a similar \teff \citep{2011A&A...535A..22C}. 
The second issue of lines being too narrow, points to velocities or velocity dispersions that are too low in the line forming region (Fig.~\ref{fig-spec3Dzoom}). In cooler RSG simulations convective velocities tend to diminish, which would amplify the discrepancy. We don't know yet if this problem is alleviated in non-grey simulations. 

\subsection{Stellar parameters : the difficulty with \teff}

The spectral energy distribution (SED) of the non-grey model described above cannot be reproduced by a single 1D hydrostatic model SED. The optical part of the spectrum resembles a cool 1D model, whereas the IR part is closer to a hotter model. Already in their study of Magellanic Cloud red supergiants, \citet{2006ApJ...645.1102L} found that \teff\ derived from optical spectrophotometry were lower than \teff\ from V-K colors. They surmised that this could result from granulation, based on a toy model combining two hydrostatic 1D MARCS models. In a very recent study based on  VLT X-shooter \citep{2011A&A...536A.105V} optical and infrared spectrophotometry, \citet{2013arXiv1302.2674D} found a mismatch between synthetic spectra from 1D MARCS models and Magellanic Cloud RSG  spectra. Using fits of TiO bands, global fits of the SED, and the IR flux method, they infer very different \teff, the TiO derived \teff\ always being much lower. In order to test the idea that the large scale and high contrast granulation might explain these differences, \citet{2013arXiv1302.2674D} fit the SED computed by \citet{2011A&A...535A..22C} for a single non-grey snapshot with 1D MARCS SEDs. It turns out that the result is very similar to the fit of real star spectra with MARCS SEDs : the global shape of the SED is reproduced by the MARCS model spectrum, specially in the infrared, but the TiO bands are stronger in the 3D model spectrum, just as seen in the observations. It thus seems that the discrepancy between red supergiant spectra and MARCs model spectra, and the offset  between \teff\ derived from TiO bands and IR spectrophotometry might be explained by  the presence of large-scale granulation. There is, however, at least one remaining question. The integrated luminosity of the 3D snapshot is 89\,000\,L$_{\odot}$, whereas the MARCS best fitting model has \teff=3700K. Using the radius $R\approx 850\, {\rm R_{\odot}}$\footnote{The radius of the 3D snapshot is defined on a radially averaged $<$3D$>$ structure, as the layer where $T=$ \teff. The temperature gradient in this region is large enough for the value of R not to depend very much on the choice of \teff\ for the reference layer.} of the 3D simulation leads to $L=4\pi R^2\sigma T_{\rm eff}^4 = 121\,000$. This 35\% difference in luminosity is most probably caused by the flux being radiated only by the hotter granules, with the inter-granular lanes only marginally contributing, depending on wavelength. The stellar disk cannot be called a stellar disk anymore, and if the effective temperature concept is to be used, one should think of introducing its "effective radius" counterpart.
A detailed study based on  a grid of 3D CO$^5$BOLD simulations is 
highly desirable, in order to fully understand red supergiant spectra and to use them to derive stellar parameters, in particular \teff\ and radius.

\section{Prospects}
In conclusion, 1D models show their limits in the world of RSGs, with concepts such as \teff\ requiring in depth rethinking. Radiative-hydrodynamics simulations are needed to achieve an even basic understanding of RSG atmospheres and spectra. Preliminary experiments with a single non-grey simulation indicate that we are getting closer to the real picture. We must now compute a (small) grid of such models, and use it to investigate spectra. High-resolution line profile studies will allow to check the predicted velocity fields, whereas well calibrated spectrophotometry will give a grasp on the thermal gradients and granulation structure.
The computational cost is however considerable, especially as models have to be run at higher spatial resolution, and with more opacity bins, until we are sure that no major physical effect is left out. The inclusion of radiative pressure may  also affect RSG simulations. The post-processing of the simulations and the computation of detailed spectra is also a tremendous effort, especially if the LTE approximation is to be lifted. In LTE, though, the Optim3D code coupled with Turbospectrum for the tabulation of opacities, comes as a handy and efficient tool to compute large chunks of spectra. Future developments include the implementation of scattering.

\begin{acknowledgements}
We wish to warmly thank the whole CO$^5$BOLD development team, who allowed us to embark on this 3D adventure into a vastly unknown territory. The reward makes the time and effort worthwhile.
Bernd Freytag is particularly thanked for enlightening discussions through the years.
We are grateful to the organizers for a very nice stay in Heidelberg and for a stimulating workshop.
The french CNRS Programme National de Physique Stellaire provided partial financial support.
\end{acknowledgements}

\bibliographystyle{aa}
\bibliography{mem_co5bold-plez-chiavassa}

\end{document}